\documentclass[aip,apl,preprint,showpacs,superscriptaddress]{revtex4-1}% for review and submission
\usepackage{graphicx}  % needed for figures
\usepackage{dcolumn}   % needed for some tables
\usepackage{bm}        % for math
\usepackage{amssymb}   % for math

\begin{document}
\title{Atomic scale electron vortices for nanoresearch}
\author{J. Verbeeck}\affiliation{EMAT, University of Antwerp, Groenenborgerlaan 171, 2020 Antwerp, Belgium}
\author{P. Schattschneider}\affiliation{Institute for Solid State Physics, Vienna University of Technology, Wiedner Hauptstrasse 8-10, A-1040 Wien, Austria}
\author{S. Lazar}\affiliation{FEI Electron Optics, 5600 KA Eindhoven, The Netherlands}
\affiliation{Canadian Centre for Electron Microscopy and Department of Materials Science and Engineering, McMaster University, Main Street West, Hamilton Ontario, Canada L8S4M1}
\author{M. St\"oger-Pollach}\affiliation{USTEM,  Vienna University of Technology, Wiedner Hauptstrasse 8-10, A-1040 Wien, Austria}
\author{S. L\"offler}\affiliation{Institute for Solid State Physics, Vienna University of Technology, Wiedner Hauptstrasse 8-10, A-1040 Wien, Austria}
\author{A. Steiger-Thirsfeld}\affiliation{USTEM,  Vienna University of Technology, Wiedner Hauptstrasse 8-10, A-1040 Wien, Austria}
\author{G. Van Tendeloo}\affiliation{EMAT, University of Antwerp, Groenenborgerlaan 171, 2020 Antwerp, Belgium}

\date{\today}

\begin{abstract}
Electron vortex beams were only recently discovered and their potential as a probe for magnetism in materials was shown. Here we demonstrate a new method to produce electron vortex beams with a diameter of less than 1.2 \AA. This unique way to prepare free electrons to a state resembling atomic orbitals is fascinating from a fundamental physics point of view and opens the road for magnetic mapping with atomic resolution in an electron microscope.
\end{abstract}

\maketitle

\begin{figure}
\includegraphics[width=0.6\columnwidth]{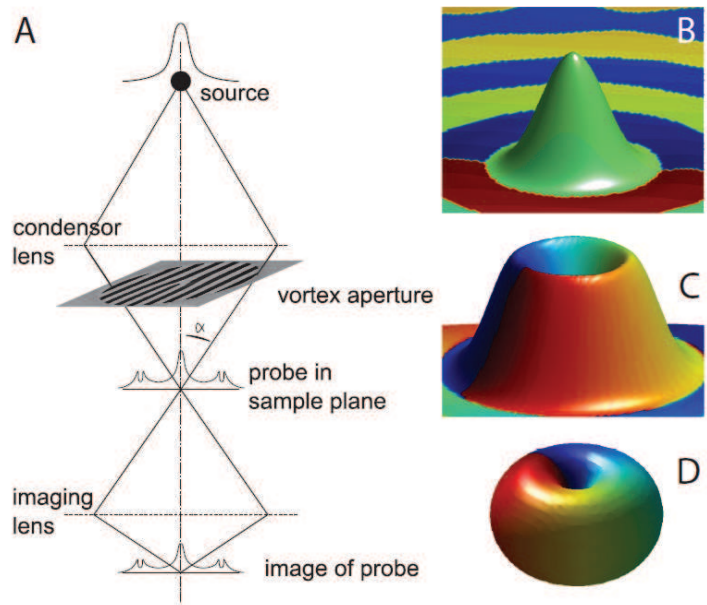}
\caption{\label{fig:1} A) Sketch of the setup to create focussed vortex probes in a transmission electron microscope. The probe is formed in the sample plane and can be used to perform atomic resolution experiments in that plane. The probe is magnified for observation by the imaging system. The convergence angle $\alpha$ can be adjusted which allows to tune the size of the vortex. B,C) Artist impression of the intensity distribution for a conventional Airy disc and a vortex beam with the same opening angle. D) Sketch of the surface of a $2p_{1}$ orbital in nitrogen containing 80\% of the electron density. The image is approximately to scale with B,C for our experimental setting of $\alpha=21.4~mrad$. Color coding indicates the phase distribution from 0 (blue) to $2\pi$ (red). Note the big simmilarity in both phase and spatial distribution between an angstrom vortex probe and an atomic orbital of a light element as nitrogen.}
\end{figure}

%\section{\label{sec:level1}First-level heading}
% sections are not used for PRL papers
The first experimental realisation of laser light carrying topological charge came in 1990 \cite{Bazhenov}, founded on a theory of field vortices \cite{Nye}. Optical vortices, as they are called, have opened a new era in optics \cite{Allen,Heckenberg}. Today, there are many applications ranging from optical tweezers exerting a torque \cite{He} over optical micromotors \cite{Luo,Friese,Franke}, cooling mechanisms \cite{Kuppens}, toroidal Bose Einstein condensates \cite{Tsurumi}, exoplanet detection \cite{Foo,Swartzlander,Berkhout,Serabyn} to quantum correlation and entanglement in many-state systems \cite{Mair,Pors,Barreiro,Leach}. For a review, see \cite{Molina}. However, all these methods are limited to the $\mu$m scale by the wavelength of light. Recently, researchers ventured to break the barrier to the atomic world: After the observation of electrons with helical wave fronts \cite{Uchida}, Verbeeck et al. successfully engineered electron vortex beams in the transmission electron microscope (TEM) \cite{Verbeeck} and the feasibility of reliably producing electron vortices with topological charge was demonstrated \cite{Verbeeck,McMorran}. With an effective diameter of several micrometers, these were still far away from the goal of atomic resolution.\\
Here, we describe the production of free electrons localised in Angstrom sized regions and carrying topological charge. Electron vortex beams are free electrons carrying a discrete orbital angular momentum of m$\hbar$. They are characterized by a spiraling wavefront, similar to optical vortices \cite{Allen}. They also carry a magnetic moment, even for beams without spin polarization equal to one Bohr magneton per electron and per unit of topological charge \cite{Bliokh,PStheory}. Their original interest was in the use as a filter for magnetic transitions \cite{Verbeeck}, thus facilitating energy loss magnetic chiral dichroism (EMCD) experiments in the electron microscope \cite{PSEMCD,Rubino,PSrealspace}. Their actual potential is much wider, ranging from probing chiral structures to the manipulation of nanoparticles, clusters and molecules, exploiting the transfer of angular momentum and the magnetic interaction.
The holographic aperture used in \cite{Verbeeck} was located in a position conjugate to the object plane. Here we use an aperture in the condensor plane of a TEM. This setup allows to form a small probe in the object plane of the microscope by means of the condensor lenses as schematically outlined in fig.\ref{fig:1}A. For an ideal electron point source and ideal lenses, the probe is given by the Fourier transform of the aperture \cite{Verbeeck}. In practice however, there are two effects which have to be taken into account. First of all, the electron source is not a point emitter but can be modelled as an incoherent source distribution over an area characterising the source size. The effect on the probe is a convolution of the intensity of the image produced by a point source with this distribution. 
A second effect, affecting the probe size is caused by the aberrations of the probe forming lens system. These can be expressed as a distortion of the ideal wavefront by a phase change.
 
A goal in the design of electron microscopes is to minimise both effects as much as possible. Current state of the art electron microscopes can reach probe sizes which deliver a resolution of better than 0.8~\AA~when used in a scanning probe approach\cite{Erni}. Here we use such a state of the art microscope to produce electron vortex beams making use of a holographic mask. The microscope used is the Qu-Ant-EM microscope installed at the university of Antwerp. This is a double aberration corrected FEI Titan G2 80-300 instrument capable of routinely making small probes which enable 0.8~\AA~resolution at an acceleration voltage of 300~kV.

Imaging such a fine  probe requires a second set of lenses with similar requirements as the probe forming lenses. Therefore, another aberration correction device in the image formation lens system is used. Nevertheless, no imaging lens is perfect and the image obtained will always \emph{overestimate} the real size of the probe. Chromatic aberration due to a finite energy spread in the gun and image blurring in the electron detector further increase the size of the \emph{image} of the probe.

The convergence semi angle $\alpha$ can be changed over a wide range which enables the user to choose between very small probes (large convergence angles) or larger probe size (smaller convergence angles). As a probe defining aperture we use a similar holographic mask with a fork dislocation as described in \cite{Verbeeck} but now with a diameter of 50 $\mu$m. Details on aperture manufacturing are given in supplementary information\cite{supp}.

\begin{figure}
\includegraphics[width=0.6\columnwidth]{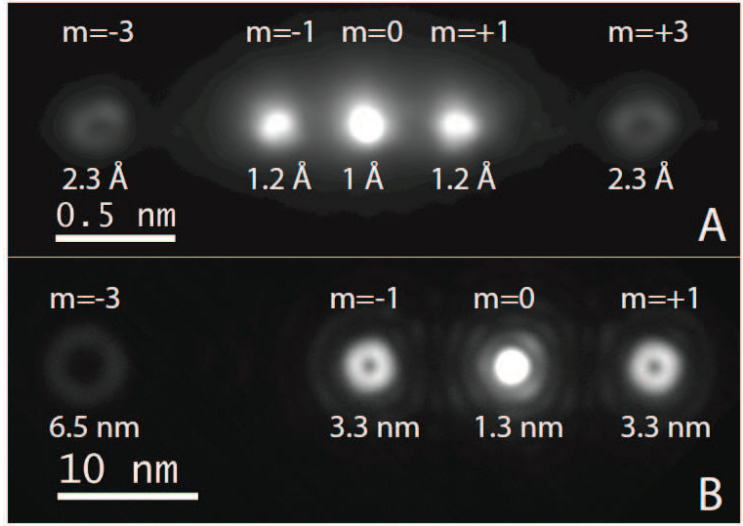}
\caption{\label{fig:2} A) Experimental probe image for a convergence angle of 21.4~mrad showing the main sidebands carrying opposite angular momentum of $\pm m \hbar$ and with a full width at half maximum of 1.2~\AA. The doughnut shape is nicely visible in the $m=\pm3$ sidebands but no minimum is observed in the $m=\pm1$ sidebands. This is due to a finite source size distribution. B) At a convergence angle of 0.8~mrad a fully diffraction limited situation is obtained where aberrations and source size effects play no significant role.}
\end{figure}

Choosing a convergence semi angle of 21.4~mrad (typically used to obtain a resolution of 0.8~\AA) we obtain the intensity distribution in fig.\ref{fig:2}A. We observe two main sidebands similar to the ones shown in our previous work but now with a full width at half maximum (FWHM) diameter of approximately 1.2~\AA. The central beam on the other hand is even smaller with a FWHM of under 1.0~\AA. The 3rd order sideband is also faintly visible with a FWHM of 2.3~\AA. Note that the first order sidebands do not show the typical doughnut shaped intensity profile which are a signature of vortex beams. The reason for this is that at such small probe sizes, the source distribution becomes dominant and smears out the intensity of the doughnut shaped beams. This effect of an extended incoherent source was discussed in detail in optics \cite{Swartzlander2007} and leads to a degradation of the vortex character of the beam in the center while the vorticity is preserved further away from the center. Intensity profiles are compared to simulated intensity profiles in supplementary information taking into acount a Gaussian source size distribution of FWHM 0.7~\AA\cite{supp}. This shows that, under these conditions, the small probes are dominated by the effect of finite source size. Without this source size effect an $m=\pm 1$ vortex would have a FWHM diameter as low as 1~\AA~which can be lowered further by increasing the convergence semi angle $\alpha$.

Reducing the convergence semi angle to 0.8~mrad leads to nanometer sized probes but now the shape is nearly perfect since the lens aberrations and source size play almost no role at this scale as shown in fig.\ref{fig:2}B. The $m=\pm 1$ vortex has a FWHM of approximately 3.3~nm while the central $m=0$ beam has a FWHM of 1.3~nm. The $m=-3$ beam has a FWHM of 6.5~nm. Probes like this are easy to make and put only very modest demands on the performance of the electron microscope as is shown in supplementary information for a non-aberration corrected FEI Tecnai F20 80-200 with an acceleration voltage of 200~kV and a convergence semi-angle of $\alpha=6.9$~mrad leading to an $m=0$ probe of 2.4~\AA~ and $m=\pm1$ probe of 4.6~\AA\cite{supp}. 

Our experiments demonstrate that it is possible to obtain sub-nm free electron vortices even on standard equipment. Aberration correctors allow vortices with diameters as small as 1.2~\AA. Comparing this to the size of a typical orbital in atoms, the shape of the beam in the focal plane resembles the electron distribution of e.g. a $2p$-orbital in a Nitrogen atom as sketched in fig.\ref{fig:1}B,C,D in both radial distribution and phase. The main difference between electrons in an atomic orbital and the free vortex is that in the latter, the wave function evolves in time as the electron propagates in the electron optical system. A detailed theoretical background of the properties of free electrons in a vortex state is given in \cite{PStheory}.
The finite source size of the electron gun is not detrimental to the typical doughnut shape of a vortex down to the sub-nm scale but limits the ability to observe the smallest Angstrom sized vortex beams that can be produced for the time being. At the same time, the finite source size, reduces the vortex character of the beam in the center while maintaining its characteristics further away from the optical axis as was studied in optics \cite{Swartzlander2007}. This would lead to a reduction of any scattering effect that hinges on the vorticity of the probe. Ongoing simulations show the trend that useful effects in inelastic scattering remain as long as the finite source size is smaller than the difraction limited size of the beam that would be obtained with an ideal point source\cite{verbeeckeelstheory}. Nevertheless, a further reduction of the finite source size in future electron microscopes would be strongly desirable for vortex experiments.

Engineering these atomic sized electron vortices opens the road to magnetic information mapping on the atomic scale \cite{PSFe}. Indeed it was shown in \cite{Verbeeck} that electron vortex beams provide information on the magnetic state of materials. With Angstrom sized electron vortices, one would obtain magnetic information on the atomic scale. In this paper we have measured the diameter of such vortex probes as their full width at half maximum, the resolution that can be obtained with such a probe is better for two reasons. First, as already mentioned, the measurement we present here is an overestimate, and source size effects play an important role. Secondly, resolution in electron microscopy is commonly defined as the spatial frequency that still gives an interpretable contrast. This difference is apparent from the measurement of the central beam which was found to have a FWHM of 1.0~\AA~while the resolution which can be obtained with this probe is approximately 0.8~\AA. Extrapolating this to the sidebeams that carry angular momentum we could estimate the possible resolution to be less than 1~\AA.
Sub-nm free electrons with topological charge can be produced in standard TEM equipment. Aberration correctors allow vortex probe sizes of less than 1.2~\AA. The dominant factor that puts a limit on the probe size is the finite electron source size. The probe with topological charge $m$ focussed on the specimen has a phase structure, extension and radial intensity distribution very similar to atomic p-orbitals even in light atoms. This fact opens new options to couple a fast electron probe directly to the internal degrees of freedom of atoms and allows to probe magnetic information on a sub nm level.
 
\section{Acknowledgements}
This work was supported by funding from the European Research Council under the 7th Framework Program (FP7), ERC grant N°246791 - COUNTATOMS. J. V. acknowledges funding from the European Research Council under the 7th Framework Program (FP7), ERC Starting Grant 278510 VORTEX. The Qu-Ant-EM microscope was partly funded by the Hercules fund from the Flemish Governement. P.S. acknowledges financial support of the Austrian Science Fund (FWF): I-543-N20. M.S-P. and A.S-T. thank J. Hell for production of the mask substrate.

\end{document}